# Analysis of a HAPS-Aided GNSS in Urban Areas using a RAIM Algorithm

Hongzhao Zheng, *Member, IEEE*, Mohamed Atia, *Senior Member, IEEE,* and Halim Yanikomeroglu, *Fellow, IEEE*

*Abstract*—The global averaged civilian positioning accuracy is still at meter level for all existing Global Navigation Satellite Systems (GNSSs), and the performance is even worse in urban areas. At lower altitudes than satellites, high altitude platform stations (HAPS) offer several benefits, such as lower latency, less pathloss, and likely smaller overall estimation error for the parameters associated in the pseudorange equation. HAPS can support GNSSs in many ways, and in this paper we treat the HAPS as another type of ranging source. In so doing, we examine the positioning performance of a HAPS-aided GPS system in an urban area using both a simulation and physical experiment. The HAPS measurements are unavailable today; therefore, they are modeled in a rather simple but logical manner in both the simulation and physical experiment. We show that the HAPS can improve the horizontal dilution of precision (HDOP), the vertical dilution of precision (VDOP), and the 3D positioning accuracy of GPS in both suburban and dense urban areas. We also demonstrate the applicability of a RAIM algorithm for the HAPS-aided GPS system, especially in the dense urban area.

## I. INTRODUCTION

TODAY, many countries and the European union have their own global navigation satellite systems (GNSSs). However, 95 percent of the time, the global averaged horizontal positioning accuracy of existing GNSSs is still at the meter level, and it is even worse for the vertical positioning accuracy [1]-[4] due to the nature of the satellite geometry. Although vertical positioning performance is less important than horizontal positioning performance today, it might be very important in the future, for instance, for unmanned aerial vehicles (UAVs) flying in the 3D aerial highways [5]. Thanks to ongoing research on localization and navigation fields, there are a number of techniques developed which can bring the positioning accuracy of systems involving satellites to the centimeter level. For example, Li et al. have shown that centimeter-level positioning accuracy can be achieved using the multi-constellation GNSS consisting of Beidou, Galileo, GLONASS and GPS with precise point positioning (PPP) [6]. Because most civilian applications use single-frequency, low-cost receivers for localization and navigation, many advanced positioning algorithms, including PPP that delivers centimeter level positioning accuracy, cannot be implemented. Therefore, the single point positioning (SPP) is the most commonly used algorithm in civilian applications. But this is poised to change. As increasing numbers of low-Earth-orbit (LEO) satellites are launched into space, researchers are investigating the feasibility of utilizing LEO satellites to aid the positioning service. For instance, Li et al. have shown that a centimeter level Signal-In-Space Ranging Error (SISRE) in the real-time PPP application can be achieved using a LEO enhanced GNSS [7]. In the event that GNSS signals are unavailable in urban areas, researchers are also interested in building navigation systems that exclusively rely on LEO satellite signals. For example, a position root mean squared error (RMSE) of 14.8 m for a UAV has been proven feasible using only two Orbcomm LEO satellites with the carrier phase differential algorithm [8]. Compared to medium-Earth-orbit (MEO) satellites, which are typically used in GNSSs, LEO satellites offer several advantages, such as lower latency and less pathloss due to shorter distance to ground users. LEO satellites also offer greater availability due to the large number of them.

To further enhance high bandwidth networking coverage in areas with obstacles, such as urban areas, another option is the use of high altitude platform stations (HAPS[1]), which refer to aerial platforms positioned in the stratosphere with a typical altitude of about 20 km. HAPS can be utilized for many technologies coming in 5G even 6G and beyond such as computation offloading [9], edge computing [10], and aerial base station [11]. As urban areas are where GNSS positioning performance degrades severely, while also being where most people live, we could improve the positioning performance of GNSS by placing several HAPS above metro cities and equipping them with satellite-grade atomic clocks so that HAPS can be deployed as another type of ranging source. Even though atomic clocks on satellites are highly accurate, they are not perfect due to the time dilation postulates made in both Einstein's special theory of relativity and the general theory of relativity. According to Einstein's special theory of relativity, an atomic clock on a fast-moving satellite runs slower than a clock on Earth by around 7 microseconds per day. On the other hand, according to the general theory of relativity, an atomic clock which experiences weaker gravity on a distant satellite runs faster than a clock which experiences greater gravity on

---

This paper was supported in part by Huawei Canada.

H. Zheng, M. Atia, and H. Yanikomeroglu are with the Department of Systems and Computer Engineering, Carleton University, Ottawa, ON K1S 5B6, Canada (e-mail: hongzhaozheng@cmail.carleton.ca; Mohamed.Atia@carleton.ca; halim@sce.carleton.ca).

[1] In this paper, the acronym "HAPS" is used to denote "high altitude platforms station" in both singular and plural forms, in line with the convention adopted in the ITU (International Telecommunications Union) documents.

Earth by about 45 microseconds per day [12]. As HAPS operate at an altitude of around 20 km and can be quasi-stationary, the time dilation is negligible from the perspective of special relativity and greatly reduced from the perspective of general relativity. Therefore, the atomic clocks on HAPS will likely be more accurate than that on satellites, which can make the estimation error of the HAPS clock offset smaller than that of the satellite clock offset. Since HAPS are positioned much closer to the Earth than satellites, the pathloss of a HAPS is expected to be much less, which will likely make the received signal power of a HAPS stronger than that of a satellite, thereby reducing the estimation error of the parameters associated in the pseudorange measurement of the HAPS signal. The movement of a HAPS can be confined to a cylindrical region with a radius of 400 m and a height of about 700 m [13], which can reduce the number of handovers during the course of navigation and increase the utilization efficiency during its operation life. As HAPS are positioned in the stratosphere, which is below the ionosphere, their signals will likely be free of the ionospheric effect, which is known to be one of the major sources of error in pseudorange measurements. Therefore, the overall estimation error for a HAPS will likely be smaller than that of a satellite. Similar to the pseudorange measurement for a satellite, which incorporates the satellite position error, we should also consider the position error in the pseudorange measurement for HAPS. Fortunately, researchers have been investigating the positioning of HAPS and have demonstrated that HAPS positioning errors are comparable to or lesser than satellite orbit errors. For example, Dovis et al. prove that 0.5 m positioning accuracy (circular error probable [CEP] 68 percent) for a HAPS is achievable using the modified RTK method [14]. There are a handful of papers in the literature that have investigated the HAPS-aided GNSS [15]-[18]; however, to the best of our knowledge, this paper is the first to provide a comprehensive study of the positioning performance of a HAPS-aided GNSS in urban areas.

There are plenty of operational GPS satellites that could fail due to the degraded signal quality for reasons such as obstruction, multipath, intentional or unintentional attacks, thereby impacting the positioning performance of the GNSS. In this case, a signal selection algorithm like the receiver autonomous integrity monitoring (RAIM) algorithm, which can detect and exclude poor quality signals, can be helpful in improving the positioning performance. For example, about 35 percent decrease in RMS positioning error of the GPS-only case and 50 percent decrease in RMS positioning error of the GPS/GLONASS case in a severe urban scenario have been achieved on smartphone GNSS chips by using a RAIM algorithm [19]. Moreover, Yang and Xu propose a robust estimation-based RAIM algorithm that can detect and exclude multiple faulty satellites effectively with efficiency higher than the conventional least squares (LS)-based RAIM algorithm [20]. In this paper, we make three postulations: 1) a HAPS signal is free of the ionospheric effect; 2) the estimation error of the HAPS clock offset is smaller than that of the satellite clock offset; and 3) the received signal power of a HAPS is higher than that of a satellite, all of which contribute toward the assumption that the overall estimation error of the parameters associated in the pseudorange equation for the HAPS is smaller than that for the satellite. Under this assumption, we use the SPP algorithm developed in our prior work [21] to show that HAPS can indeed improve the positioning performance of legacy GNSSs in urban areas through both a simulation and a physical experiment. We also demonstrate the applicability of the RAIM algorithm to a HAPS-aided GPS system, especially in dense urban areas. Since the HAPS measurements are unavailable so far, they are simulated in a rather simple but logical way in both the simulation and physical experiment. The contributions of this paper are listed below.

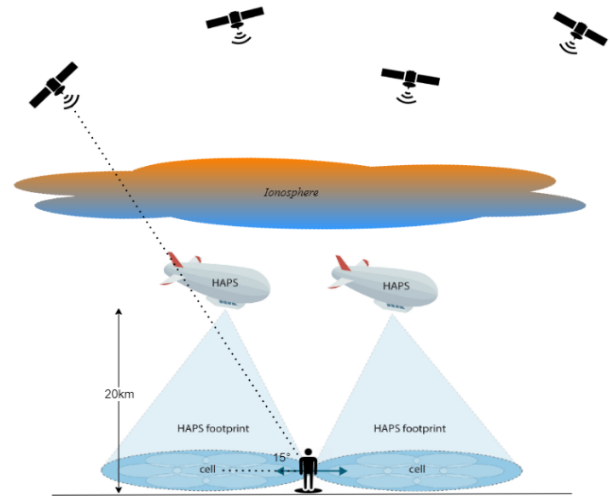

Fig. 1. System model of the HAPS-aided GPS.

- First, using a commercial GNSS simulator, we simulate the GPS pseudorange signals and generate the positions of HAPS. By using the default system parameters as well as a proper manipulation of the number of visible satellites, we show that the positioning performance of the GPS-only system in both the suburban and dense urban areas are close to the real scenario. Moreover, we show the positioning performance of different systems where different numbers of HAPS are used with or without the GPS system. The issue of the ranging source geometry is revealed from the simulation results.
- Next, we apply the SPP algorithm to the real GPS data collected using two commercial GNSS receivers as well as the HAPS data generated using the commercial GNSS software. In so doing, we show the advantage of the HAPS-aided GPS system in the sense of the horizontal dilution of precision (HDOP) and the vertical dilution of precision (VDOP).
- Finally, we implement a RAIM algorithm and demonstrate its effectiveness in improving the 3D positioning performance of the HAPS-aided GPS system, especially in dense urban areas.

The rest of the paper is organized as follows: in Section II, the system model, the SPP algorithm, and the RAIM algorithm are described. In Section III, the simulation setup of the HAPS-aided GPS system and the simulation results are presented. In Section IV, the physical experiment setup and results, including both the DOP analysis and the 3D positioning accuracy analysis, are provided. Finally, Section V offers some conclusions and a discussion of future research directions. For simplicity, the GNSS signal only involves the GPS C/A L1 signal.

## II. System Model

The system model of the HAPS-aided GPS system is depicted in Fig. 1. There are four satellites shown in Fig. 1, this is just a reminder that at least four satellites are required to perform precise 3D localization using GNSS. The typical choice for the elevation mask is 10 degrees. However, we use 15 degrees as the elevation mask for the satellites and HAPS due to the following reasons: 1) the atmospheric error owing to the signal refraction can be neglected if the elevation of a satellite is greater than 15 degrees [22], which is likely true for a HAPS as well; 2) As there is a higher chance of ensuring the required number of ranging source with HAPS, we can improve the positioning performance further by only using those satellite signals with better quality. The pseudorange equation for satellite is given by

$$p_{SAT} = \rho_{SAT} + d_{SAT} + c(dt - dT_{SAT}) + d_{ion,SAT} + d_{trop,SAT} + \epsilon_{mp,SAT} + \epsilon_p \quad (1)$$

where $p_{SAT}$ denotes the satellite pseudorange measurement, $\rho_{SAT}$ is the geometric range between the satellite and receiver, $d_{SAT}$ represents the satellite orbit error, $c$ is the speed of light, $dt$ is the receiver clock offset from GPS time, $dT_{SAT}$ is the satellite clock offset from GPS time, $d_{ion,SAT}$ denotes the ionospheric delay for satellite signals, $d_{trop,SAT}$ denotes the tropospheric delay for satellite signals, $\epsilon_{mp,SAT}$ is the delay caused by the multipath for satellite signals, and $\epsilon_p$ is the delay caused by the receiver noise. The pseudorange equation for HAPS can be expressed as follows:

$$p_{HAPS} = \rho_{HAPS} + d_{HAPS} + c(dt - dT_{HAPS}) + d_{trop,HAPS} + \epsilon_{mp,HAPS} + \epsilon_p \quad (2)$$

where $p_{HAPS}$ denotes the HAPS pseudorange measurement, $\rho_{HAPS}$ represents the geometric range between the HAPS and the receiver, $d_{HAPS}$ represents the HAPS position error, $dT_{HAPS}$ is the HAPS clock offset from GPS time, $d_{trop,HAPS}$ denotes the tropospheric delay for HAPS signals, $\epsilon_{mp,HAPS}$ is the delay caused by the multipath for HAPS signals. The simulated vehicle trajectory originates at Carleton University, which is in a suburban area, and ends at Rideau Street, which is in a dense urban part of Ottawa (see Fig. 2). There are six simulated HAPS shown as transmitters on Fig. 3. As we can see, one HAPS is positioned over downtown Ottawa; the other five HAPS are positioned nearby, over populated areas and conservation areas. HAPS is quasi-stationary, meaning that it will still be moving in a variety of manners. In this work, all the HAPS are simulated to be following a circular trajectory with a radius of 300 m. The elevation and azimuth angles of all the HAPS at the beginning of the simulation are listed in Table I. The positions of HAPS were chosen to provide a rich diversity in azimuth angles. With one HAPS at the zenith and the others having relatively low elevation angles, this constitutes a near Zenith + Horizon (ZH) geometry, which can deliver a reasonably good DOP [23]. To make sure the entire urban area is well covered, HAPS are placed not too far away from the urban area. To better understand the concept of DOP, the visual illustrations of the HDOP and VDOP of the simulated HAPS constellation are provided in Fig. 4 and Fig. 5, respectively. Due to various errors impacting the pseudorange measurement, the estimated distance between a HAPS and a receiver can be smaller or larger than the geometric range. Objects with a higher elevation angle will likely result in more uncertainty for the vertical component and less uncertainty for the horizontal component from the point of view of geometry, and vice versa. The shaded area is where the receiver is estimated to be.

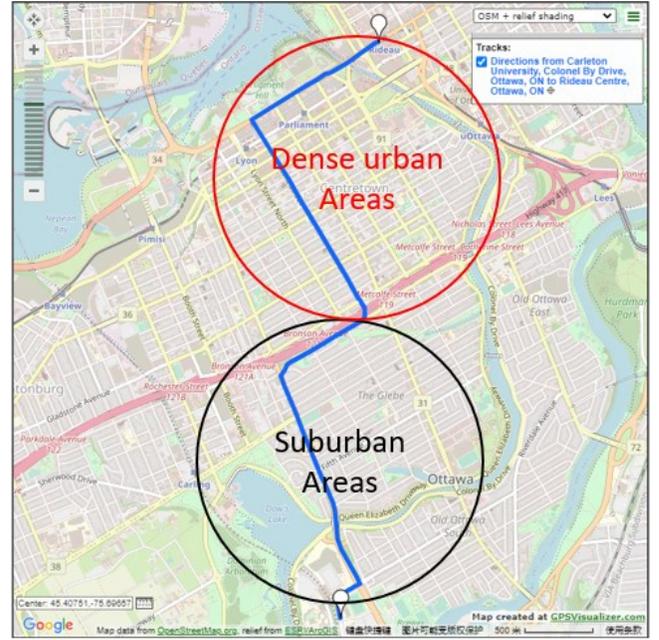

Fig. 2. Vehicle trajectory.

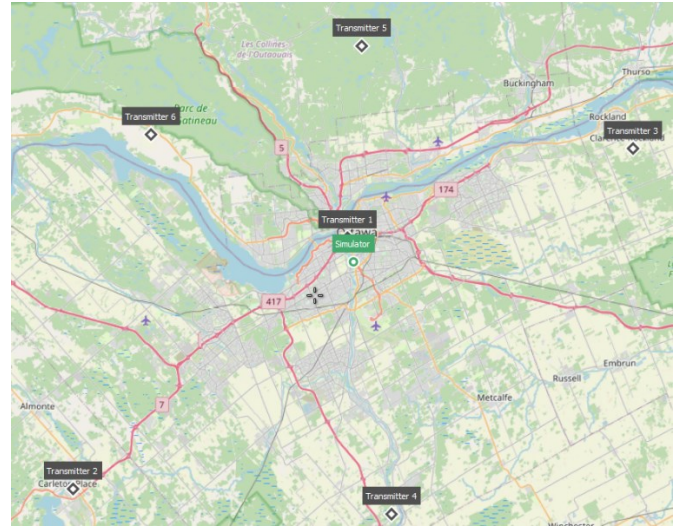

Fig. 3. Locations of the simulated HAPS.

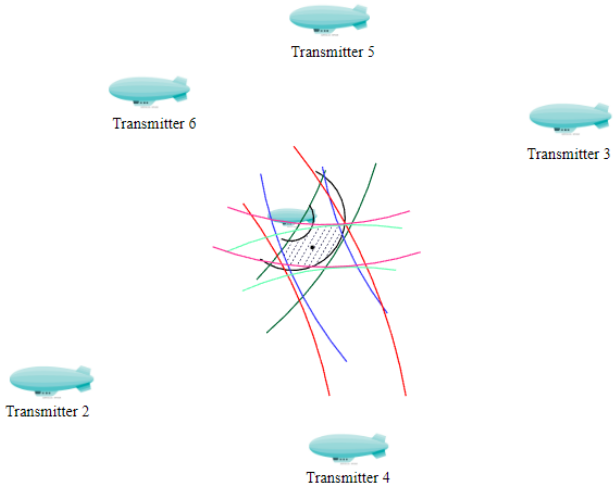

Fig. 4. HDOP of the simulated HAPS constellation (top view).

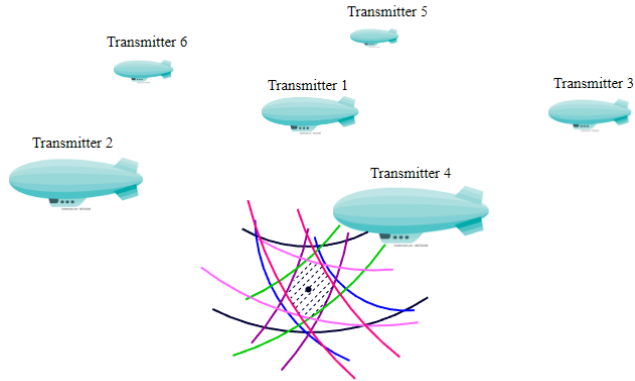

Fig. 5. VDOP of the simulated HAPS constellation (front view).

TABLE I
ELEVATION AND AZIMUTH OF THE HAPS AT THE START OF THE SIMULATION

| HAPS index | Elevation angle | Azimuth angle |
|---|---|---|
| HAPS #1 | 81.087° | -14.210° |
| HAPS #2 | 24.054° | -128.878° |
| HAPS #3 | 27.952° | 68.022° |
| HAPS #4 | 32.450° | 171.477° |
| HAPS #5 | 36.554° | 2.204° |
| HAPS #6 | 33.805° | -57.884° |

*A. The Single Point Positioning (SPP) Algorithm*

The single point positioning algorithm is implemented on the basis of the SPP package developed by Napat Tongkasem [24] with proper modifications [21] so that HAPS can be incorporated in the SPP algorithm. Fig. 6 shows the flowchart of the single point positioning algorithm. We should point out that the implemented single point positioning algorithm is not the best positioning algorithm, and that the objective of this work is to show the significance of HAPS in aiding the positioning performance of a legacy GNSS. The implemented SPP algorithm can be improved in many ways. For example, if the knowledge of the measurement error variance is available, we can apply the weighted least squares (WLS) algorithm to enhance the positioning performance of the SPP algorithm by lowering the weights of those observations with higher variances [25]. If the knowledge of the measurement error variance is unavailable, the computational complexity of the SPP algorithm can be reduced by imposing the Cholesky decomposition for the matrix inversion in (9) [26]. We can also use the carrier phase measurement to enhance the positioning performance of the HAPS-aided GPS system, since carrier performance of the HAPS-aided GPS system, since carrier phase measurements come with much higher precision, which usually delivers a more accurate position solution. Since the HAPS clock offset in this work is not explicitly simulated, we simply use $dT$ to denote the satellite clock offset. From the data collected by the GNSS receiver, we shall obtain both the receiver independent exchange (RINEX) format observation file and the RINEX navigation file, from which we can obtain satellite information, such as the satellite pseudorange $p_{SAT}$, the ionospheric parameters $\alpha$, the Keplerian parameters, the pseudo-random noise ($PRN$) code, which represents the unique number of each satellite, the day of year ($DOY$) which represents the day of year at the time of measurement. We write $PRN$ in bold to represent a vector containing the pseudo-random noise code of all visible satellites at the current epoch. We are able to compute the satellite positions, $P_{SAT}$, and satellite clock offset, $dT$, using the Keplerian parameters contained in the navigation file. $P_{HAPS}$ denotes a vector containing the positions of all HAPS, which are generated using the Skydel GNSS simulator [27], and $p_{HAPS}$ denotes a vector containing the HAPS pseudorange, which will be explained in Section III. To compute the position solution $x$, we first initialize the receiver position to the center of the Earth; then we initialize the receiver clock offset to zero and the change in estimates $dx$ to infinity. For each epoch of measurement, we first check if the number of available ranging sources is more than three, as at least four ranging sources are required to perform precise 3D localization. Since the receiver position is iteratively estimated, we calculate the elevation angles for both satellites and HAPS with respect to the recently estimated receiver position. Since both the tropospheric delay and the ionospheric delay are functions of the receiver position, these two atmospheric delays are estimated iteratively as well. The elevation angle, satellite pseudorange, HAPS pseudorange, satellite position, satellite clock offset, tropospheric delay $d_{trop}$, ionospheric delay $d_{ion}$, and pseudo-random noise ($PRN$) code are modified iteratively on the basis of the re-computed elevation angles for both satellites and HAPS. To prepare the parameters needed for the least square method, the pseudorange needs to be corrected as follows:

$$p^c_{SAT} = p_{SAT} + c \cdot dT - d_{trop,SAT} - d_{ion,SAT} \quad (3)$$

where $p^c_{SAT}$ represents the corrected pseudorange for the satellites, and $p_{SAT}$ represents the uncorrected pseudorange for the satellites. In this work, the HAPS pseudorange is modeled as the sum of the geometric range and the pseudorange error, which represents the overall estimation error of the parameters in the HAPS pseudorange equation. Accordingly, the HAPS pseudorange does not need to be corrected. Due to the Earth's rotation, the positions of satellites and HAPS at the signal emission time are different from their positions at the signal reception time; this is known as the Sagnac effect [28]. The coordinates of satellites/HAPS can be transformed from the signal emission time to the signal reception time by [28]

$$\Delta t_{ROT} = t_{rx} - t_{tx} \qquad (4)$$

$$P_{i,rx} = M_{ROT}(\omega_E \times \Delta t_{ROT}) P_{i,tx} \qquad (5)$$

where $\Delta t_{ROT}$ denotes the signal propagation time, $t_{rx}$ represents the signal reception time, $t_{tx}$ represents the signal emission time, $P_{i,rx}$ is the $i^{th}$ satellite/HAPS coordinates at the signal reception time, $P_{i,tx}$ is the $i^{th}$ satellite/HAPS coordinates at the signal emission time, $\omega_E$ denotes the Earth's rotation rate, and $M_{ROT}(\omega_E \times \Delta t_{ROT})$ is known as the rotation matrix, which is described as follows:

$$M_{ROT}(\omega_E \times \Delta t_{ROT})
= \begin{bmatrix} \cos(\omega_E \times \Delta t_{ROT}) & \sin(\omega_E \times \Delta t_{ROT}) & 0 \\ -\sin(\omega_E \times \Delta t_{ROT}) & \cos(\omega_E \times \Delta t_{ROT}) & 0 \\ 0 & 0 & 1 \end{bmatrix}. \qquad (6)$$

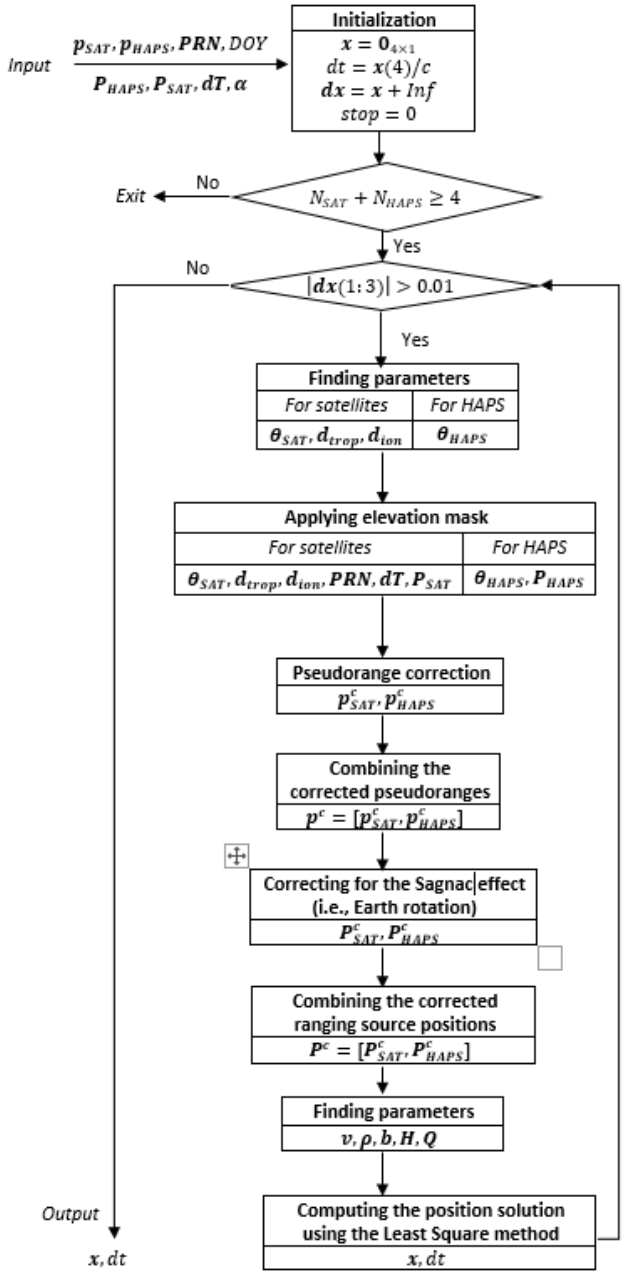

Fig. 6. Flow chart of the single point positioning algorithm.

The line-of-sight vector $v$, and the true range between ranging sources and receiver $\rho$, are then calculated to compute the a priori range residual vector $b$ and the design matrix $H$, where

$$b = p^c - \rho \qquad (7)$$

$$H = [v, \mathbf{1}_{length(P^c) \times 1}] \qquad (8)$$

where $p^c$ is the corrected satellite pseudorange combined with the corrected HAPS pseudorange, $\mathbf{1}_{length(P^c) \times 1}$ denotes a column vector of length being the total number of visible ranging sources, and $P^c$ is a vector containing the corrected positions of the visible ranging sources (i.e., satellite + HAPS). Finally, the least square solution is computed as follows:

$$Q = (H'H)^{-1} \qquad (9)$$

$$dx = QH'b \qquad (10)$$

$$dt = dx(4)/c \qquad (11)$$

where $Q$ is known as the covariance matrix, and $dx(4)$ denotes the fourth element in the vector $dx$. To prevent the algorithm from getting numerical issues, we should ensure the term being inversed in (9) is non-singular; in other words, the design matrix $H$ should be non-singular. With the extra observations by utilizing HAPS as additional ranging sources, the chance of $H$ being singular is likely reduced; the non-singular design matrix can be ensured by avoiding the use of collinear observations, which means that two or more observations have about the same azimuth and elevation angle. We observe that the term being inversed in (9), $H'H$, is a Hermitian, positive definite matrix, therefore the Cholesky decomposition can be imposed to reduce the computational complexity [26]. The covariance matrix, $Q$, is described by

$$Q = \begin{bmatrix} \sigma_x^2 & \sigma_{xy} & \sigma_{xz} & \sigma_{xt} \\ \sigma_{xy} & \sigma_y^2 & \sigma_{yz} & \sigma_{yt} \\ \sigma_{xz} & \sigma_{yz} & \sigma_z^2 & \sigma_{zt} \\ \sigma_{xt} & \sigma_{yt} & \sigma_{zt} & \sigma_t^2 \end{bmatrix} \qquad (12)$$

where receiver coordinates x, y, z in the Earth-centered Earth-fixed (ECEF) coordinate frame and the receiver clock offset, respectively. The least square solution will be found when the norm of the change in receiver position $dx(1:3)$ is sufficiently small. In this work, this threshold is set as 0.01 m. We use the HDOP, the VDOP and the 3D positioning accuracy as the metrics to show the advantage of the proposed HAPS-aided GPS system; the 3D positioning accuracy is used to show the applicability of the RAIM to the HAPS-aided GPS system. To compute the HDOP, we need to convert the covariance matrix into the local north-east-down (NED) coordinate frame, which can be done with the following equations [29]:

$$Q_{NED} = R'\widetilde{Q}R = \begin{bmatrix} \sigma_n^2 & \sigma_{ne} & \sigma_{nd} \\ \sigma_{ne} & \sigma_e^2 & \sigma_{ed} \\ \sigma_{nd} & \sigma_{ed} & \sigma_d^2 \end{bmatrix} \quad (13)$$

$$\widetilde{Q} = \begin{bmatrix} \sigma_x^2 & \sigma_{xy} & \sigma_{xz} \\ \sigma_{xy} & \sigma_y^2 & \sigma_{yz} \\ \sigma_{xz} & \sigma_{yz} & \sigma_z^2 \end{bmatrix} \quad (14)$$

$$R = \begin{bmatrix} -\sin\lambda & \cos\lambda & 0 \\ -\cos\lambda\sin\varphi & -\sin\lambda\sin\varphi & \cos\varphi \\ \cos\lambda\cos\varphi & \sin\lambda\cos\varphi & \sin\varphi \end{bmatrix} \quad (15)$$

where $\sigma_n$, $\sigma_e$, and $\sigma_d$ represent the receiver position errors in the local north, east, and down directions, respectively. $\lambda$ and $\varphi$ represent the longitude and latitude of the receiver, respectively. Then, the HDOP is described by

$$HDOP = \sqrt{\sigma_n^2 + \sigma_e^2} \quad (16)$$

and the VDOP is described by

$$VDOP = \sqrt{\sigma_d^2}. \quad (17)$$

*B. The Receiver Autonomous Integrity Monitoring (RAIM) Algorithm*

The RAIM algorithm is a signal selection algorithm that can detect and even exclude abnormal observations using redundant measurements. It can detect an abnormal observation when the number of observations is at least five; it can exclude this abnormal observation when the number of observations is at least six. The RAIM algorithm is typically applied to multi-constellation GNSSs where the number of ranging sources is more than enough to perform precise 3D localization, and it is typically applied to cases where there likely exists at least one observation that differs from the expected value significantly. Such cases include urban areas, where the pseudorange measurement is highly subject to the multipath effect. With the assistance from HAPS, the chance of enabling the RAIM function will likely increase. Typical RAIM algorithms tend to use the standard deviation of the target observable, which is the pseudorange measurement in our work. As knowledge of the standard deviation of the satellite pseudorange is unavailable on the receivers we use, in this work the RAIM algorithm is implemented on the basis of [30], which considers a $C/N_0$-based variance model and a computationally efficient method, namely the modified Danish estimation method.[2] The implemented $C/N_0$-based RAIM algorithm is given in Alg. 1, where $N$ denotes the number of visible ranging sources. The input to this algorithm consists of the position fix computed using the SPP algorithm $x$, and the $C/N_0$ of the ranging source signal. Since HAPS are located at much lower altitudes than

---

**Algorithm 1** The $C/N_0$-based RAIM Algorithm

**Input**: The SPP estimated position solution $x$ and $C/N_0$;
**Output**: The SPP and RAIM jointly estimated position solution $\hat{x}$.

1: Initialize the parameters *stop* and $dx$;
2: **while** $|dx(1:3)| > 0.01$ **do**
3:     *Same procedures as the SPP algorithm until "Finding parameters" after correcting for the Sagnac effect*;
4:     **for** $i = 1$ to $N$ **do**
5:        **if** $stop == 1$ **do**
6:           Find the variance of the observation $i$, $s_i$, according to (19);
7:        **end if**
8:     **end for**
9:     Find the weight matrix $W$ and the design matrix $H$, and calculate the covariance matrix $Q$ according to (20);
10:    Calculate the change in estimates $dx$ according to (21), and update the position solution $x$;
11:    Calculate the pseudorange residual $\hat{v}$ according to (22), and the covariance matrix of the residuals $C_{\hat{v}}$ according to (24);
12:    **for** $i = 1$ to $N$ **do**
13:       Find the normalized residual of observation $i$ at the current iteration $k$, $\bar{w}_{i,k}$ according to (26);
14:       **if** $|\bar{w}_{i,k}| > n_{1-(\alpha_0/2)}$ **do**
15:          Update the variance of the observation $i$ for the next iteration $k+1$, $\sigma_{i,k+1}^2$, according to (25);
16:       **end if**
17:    **end for**
18: **end while**

---

[2] To the best of our knowledge, RAIM is the most common algorithm used for integrity monitoring. Since we only have the $C/N_0$ data which can be utilized for the integrity monitoring, we could not identify in the literature any other appropriate RAIM-like algorithm for comparison. However, we believe that the other RAIM algorithms would also be applicable if the knowledge of the standard deviation of the satellite pseudorange happens to be available.

satellites, in practice the $C/N_0$ value of the HAPS might be higher than that for any satellite. As it is possible that a handful of HAPS signals might suffer from severe multipath effects, we can exclude those HAPS signals whose $C/N_0$ values are much lower than the higher ones. In this work, the multipath effect is not explicitly simulated for the HAPS signal; therefore, we assume that the $C/N_0$ of each HAPS is equal to the maximum $C/N_0$ value of the available satellites at each epoch, meaning that the signal quality for a HAPS will always be better than that for any satellites. The variance covariance matrix (VCM) $\boldsymbol{\Sigma}$ of the observations (pseudoranges) $\boldsymbol{p}$ is defined as follows:

$$\boldsymbol{\Sigma} = diag(s_1, s_2, \dots, s_n) \tag{18}$$

$$s_i = 10 + 150^2 * 10^{(-C/N_{0,i})/10} \tag{19}$$

where $s_i$ denotes the variance of the observation $i$. We assume that the observations are uncorrelated, and that the errors follow the normal distribution with $N(\boldsymbol{0}, \boldsymbol{\Sigma})$. The weight matrix, $\boldsymbol{W}$, is defined as the inverse of the VCM, $\boldsymbol{\Sigma}^{-1}$. The least square equations become

$$\boldsymbol{Q} = (\boldsymbol{H}'\boldsymbol{W}\boldsymbol{H})^{-1} \tag{20}$$

$$\boldsymbol{dx} = \boldsymbol{Q}\boldsymbol{H}'\boldsymbol{W}\boldsymbol{P}. \tag{21}$$

The least square residuals of the pseudorange $\hat{\boldsymbol{v}}$ can be obtained as follows:

$$\hat{\boldsymbol{v}} = \boldsymbol{H} \cdot \boldsymbol{dx} - \boldsymbol{P} \tag{22}$$

$$\boldsymbol{P} = \boldsymbol{p}^c - \boldsymbol{\rho} \tag{23}$$

where $\boldsymbol{H}$ represents the design matrix, $\boldsymbol{dx}$ represents the change in estimates, $\boldsymbol{p}^c$ denotes the corrected pseudoranges, and $\boldsymbol{\rho}$ denotes the geometric range between ranging sources and the receiver. The covariance matrix of the residuals, $\boldsymbol{C}_{\hat{v}}$, is computed as

$$\boldsymbol{C}_{\hat{v}} = \boldsymbol{\Sigma} - \boldsymbol{H}(\boldsymbol{H}^T\boldsymbol{\Sigma}^{-1}\boldsymbol{H})^{-1}\boldsymbol{H}^T. \tag{24}$$

To detect and exclude the abnormal observations, we follow the modified Danish estimation method proposed in [30].

$$\sigma_{i,k+1}^2 = \sigma_{i,0}^2 \cdot \begin{cases} \exp\left(\frac{|\bar{w}_{i,k}|}{T}\right), & |\bar{w}_{i,k}| > n_{1-(\alpha_0/2)} \\ 1, & |\bar{w}_{i,k}| \le n_{1-(\alpha_0/2)} \end{cases} \tag{25}$$

with

$$\bar{w}_{i,k} = \frac{\hat{v}_{i,k}}{\sqrt{(C_{\hat{v}_{i,1}})_{ii}}} \tag{26}$$

where $\sigma_{i,0}^2$ denotes the a priori variance of the observation $i$ (i.e., $s_i$), $\bar{w}_{i,k}$ denotes the normalized residual of observation $i$ at iteration $k$, $\sqrt{(C_{\hat{v}_{i,1}})_{ii}}$ represents the standard deviation of observation $i$ from the first iteration, $n_{1-(\alpha_0/2)}$ denotes the $\alpha_0$-quantile of the standard normal distribution, which is also called the critical value, $\alpha_0$ is the predetermined false alarm rate which is 0.5 % in this work. The modified Danish method is an iteratively reweighted LS algorithm that implements a robust estimator. This method detects and excludes abnormal observations by comparing the absolute value of each normalized pseudorange residual, $|\bar{w}_{i,k}|$, with the critical value, $n_{1-(\alpha_0/2)}$, in each iteration. The variances for the observations whose normalized residuals are greater than the critical value are multiplied with exponential terms, making the variances of those observations larger, hence lowering the weight of those observations. By iteratively multiplying the variances of the abnormal observations by exponential terms, the weight of the abnormal observations will likely become much smaller than that of the normal observations; therefore, the abnormal observations can be considered as being excluded.

III. SIMULATION OF THE HAPS-AIDED GPS SYSTEM

In this section, we will describe the simulation setup used in the Skydel GNSS software [27] and present the simulation results in terms of 3D positioning accuracy for several hybrid systems and two standalone systems in both a suburban scenario and a dense urban scenario.

*A. Simulation Setup*

The system model is established using the default Earth orientation parameters of the Skydel GNSS simulation software [27], which considers all GPS satellites orbiting around the Earth and transmitting the L1 C/A code. The Saastamoinen model is chosen to emulate the tropospheric effect, and the Klobuchar model is chosen to emulate the ionospheric effect using the default Klobuchar parameters that come with the software. The output from Skydel contains the ECEF coordinates of satellites at the signal emission time, the ionospheric corrections, the tropospheric corrections, the satellite clock offsets, the ECEF coordinates of the receiver, the signal emission time, and so forth, at each time stamp from the start of the simulation. The receiver clock offset in the simulation is zero by default. The correction terms in the pseudorange equation of satellite including the satellite orbit error, the multipath error, and the receiver noise are not separately considered in the simulation; instead, a pseudorange error is introduced to reflect the presence of those effects. The pseudorange error of satellite is featured using the built-in first order Gauss-Markov process with a default time constant of 10 s, and the standard deviation of 6 m. The continuous model for the first order Gauss-Markov process is described by [31]:

$$\dot{x} = -\frac{1}{T_c}x + w \tag{27}$$

where $x$ represents a random process with zero mean, correlation time $T_c$, and noise $w$. The autocorrelation of the first

TABLE II
DETAILS OF THE SIMULATION SETUP

| Item | Processing strategy |
| --- | --- |
| Earth orientation parameter | Software default Earth orientation parameters |
| Satellite signal | GPS L1 C/A |
| Tropospheric model | Saastamoinen model |
| Ionospheric model | Klobuchar model |
| Sampling rate | 12.5 MS/s |
| Satellite pseudorange error | 1st order Gauss-Markov process (time constant = 10 s; standard deviation = 6 m) |
| HAPS pseudorange error | Gaussian noise (mean = 0 m; std = 2 m for the suburban scenario; std = 5 m for the dense urban scenario) |
| Number of GPS satellites (dense urban scenario) | 8-10 |
| Number of GPS satellites (suburban scenario) | 4 |
| Total number of HAPS | 6 |

order Gauss-Markov process is described by [32]:

$$R(\Delta t) = \sigma^2 e^{-\frac{|\Delta t|}{\tau}} \quad (28)$$

where $\Delta t$ represents the sampling interval, $\sigma$ and $\tau$ denote the standard deviation and the time constant of the first order Gauss-Markov process, respectively. The characteristics of the pseudorange errors for satellites are set to be the same in both the suburban scenario and the dense urban scenario. However, we randomly select four satellites in the dense urban scenario in order to emulate the dense urban area in a rather simple way. We verify that by doing so, the standard deviation of the 3D positioning accuracy for the GPS-only system in the simulation is close to that in the physical experiment. The pseudorange error for the HAPS is modeled using the Gaussian noise with standard deviations of 2 m and 5 m, representing the suburban and the dense urban scenario, respectively. Under the assumption that the overall estimation error of the HAPS is less than that of the satellite, the standard deviation for the HAPS pseudorange error is deliberately set to be smaller than that of the satellite pseudorange error in the suburban and dense urban scenarios. To investigate the impact of the number of HAPS on the positioning performance of the HAPS-aided GPS system, we consider four hybrid systems with different numbers of randomly selected HAPS at each epoch. We also consider the HAPS-only system for the completeness of a research problem. Under this setting, we examine the 3D positioning performance of different systems in the suburban and dense urban scenarios. In the suburban scenario, the number of visible satellites varies between eight and ten, while in the dense urban scenario the number of visible satellites is set to four. The details of the simulation setup are given in Table II.

*B. Simulation Results*

Fig. 7 shows the cumulative distribution function (CDF) of the 3D positioning accuracy for different positioning systems in the suburban scenario. With the assumption that the pseudorange error for a HAPS is smaller than that of a satellite, we can see from Fig. 7 that all the hybrid systems (HAPS +

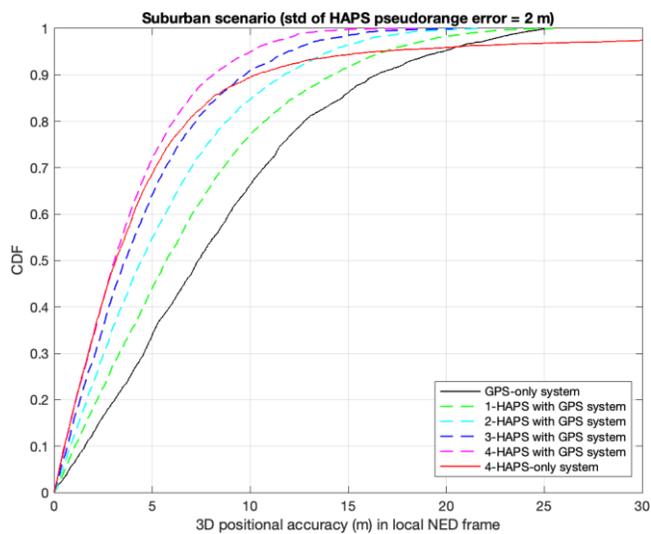

Fig. 7. CDF of the 3D positioning accuracy for different systems (suburban scenario).

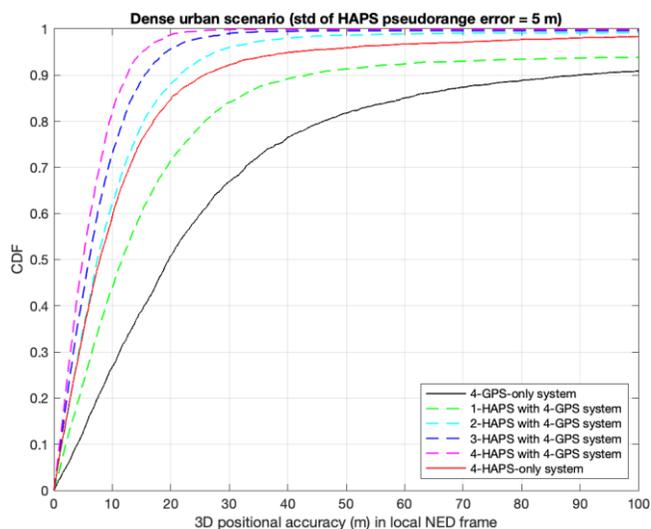

Fig. 8. CDF of the 3D positioning accuracy for different systems (dense urban scenario).

GPS) outperform the GPS-only system; the more HAPS, the better the positioning performance of the HAPS-aided GPS

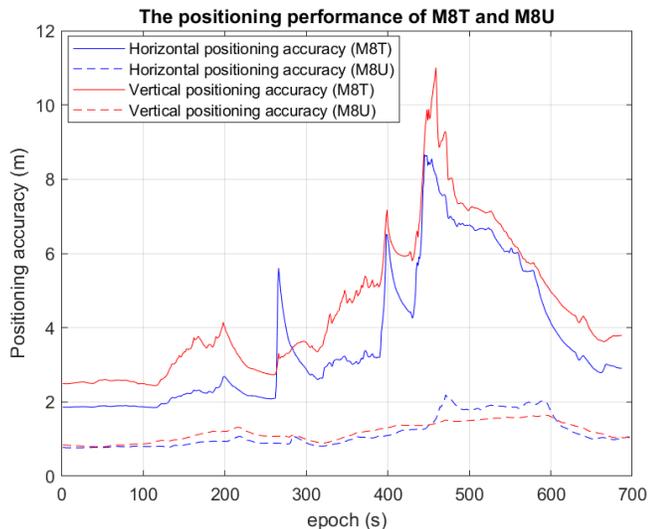

Fig. 9. Positioning performance of M8T and M8U for horizontal and vertical planes.

TABLE III
EVK-M8T GNSS UNIT SPECIFICATIONS [35]

| Parameter | Specification |
|---|---|
| Serial Interfaces | 1 USB V2.0 |
| | 1 RS232, max.baud rate 921,6 kBd DB9 +/- 12 V level 14 pin – 3.3 V logic |
| | 1 DDC (I2C compatible) max. 400 kHz |
| | 1 SPI-clock signal max. 5,5 MHz – SPI DATA max. 1 Mbit/s |
| Timing Interfaces | 2 Time-pulse outputs |
| | 1 Time-mark input |
| Dimensions | 105 × 64 × 26 mm |
| Power Supply | 5 V via USB or external powered via extra power supply pin 14 (V5_IN) 13 (GND) |
| Normal Operating Temperature | −40℃ to +65℃ |

system. Nevertheless, we observe that the positioning performance of the 4-HAPS-only system, where all ranging sources have a much smaller pseudorange error than the satellite, is not the best and can occasionally be very poor. This may be due to the following reasons: 1) the 4-HAPS-only system has much fewer ranging sources for computing receiver positions; 2) the HAPS geometry can be poor occasionally since we are randomly selecting four HAPS at each epoch. There are several cases where the HAPS geometry is considered poor. For example, when the four randomly selected HAPS are on the same side of the receiver. The CDF of the 3D positioning accuracy for different positioning systems in the dense urban scenario is shown in Fig. 8, from which we can see a similar trend that the more HAPS the better the positioning performance of the HAPS-aided GPS system. In the dense urban scenario, where only four GPS satellites are selected for positioning, the 4-HAPS-only system achieves better positioning performance than the 4-GPS-only system due to the better signal quality for the HAPS. However, we should consider using the HAPS-aided GPS system for the best positioning performance.

## IV. FIELD EXPERIMENTS

To verify and support the simulation results, we process the raw GNSS data collected using two commercial GNSS receivers. In this section, we present the experiment setup, the modeling of HAPS signals, and the HAPS pseudorange error. We also provide an analysis of the DOP and the 3D positioning accuracy for both the GPS-only system and the HAPS-aided GPS system.

### A. Experiment Setup

The raw GNSS data is collected along a vehicle trajectory similar to the one shown in Fig. 2, except for a slight difference due to a partial road closure on the day of data collection. The raw GNSS data is collected using both the Ublox EVK-M8T and the Ublox EVK-M8U GNSS units. The Ublox EVK-M8T unit is a timing product that can provide users with precise timing information for post-processing; the Ublox EVK-M8U unit is a dead reckoning product equipped with inertial measurement units (IMUs) such that the positioning performance of this product will not be degraded much even in the dense urban area. Fig. 9 shows the positioning performance along both horizontal and vertical planes for both M8T and M8U during the entire observation period. From Fig. 9, we can see that M8U outperforms M8T for both horizontal positioning accuracy and vertical positioning accuracy. As only M8T provides the timing information required for post-processing, the receiver positions computed by EVK-M8U are used as the ground truth data for the analysis of the positioning performance. The raw GNSS data is processed using the single point positioning algorithm described in Section II. Table III gives the specifications of the EVK-M8T GNSS unit. To emulate realistic LOS conditions for HAPS in the urban area, the LOS probability as a function of the HAPS elevation angle in the urban area is implemented on the basis of [33] and [34]. It is worth mentioning that the LOS probability model for the HAPS provided by [33] is proposed based on the city of Chicago; imposing this LOS probability model for the dense urban area of Ottawa might be too harsh considering their incompatible city scales. Since there is no LOS probability corresponding to the suburban area in [34], the one for rural area in [34] is used as the LOS probability for the HAPS in the suburban area. The LOS probability for the HAPS in the rural area in [34] is verified as being consistent with the LOS probability for the HAPS in the suburban area in [33]. The pseudorange of the HAPS in the experiment is modeled as the addition of the geometric range between the satellite and receiver, the receiver clock offset multiplied by the speed of light, and the pseudorange error representing the sum of all kinds of estimation errors. The pseudorange errors for the HAPS in the suburban and dense urban areas are simulated as Gaussian noise with zero mean and standard deviations of 2 m and 5 m, respectively. Since the vehicle trajectory involves both suburban and dense urban areas, the entire route is divided into two segments, where the first segment is considered as the suburban scenario while the second segment is considered as the dense urban scenario (see Fig. 2). By observing the positioning performance of the GPS-only system using the real GPS data, the LOS probability for the suburban area is applied to the HAPS for epochs less than 380 s, and the LOS probability for the dense urban area is applied to the HAPS for epochs greater than or equal to 380 s. Since the GNSS receivers we use do not provide an accurate receiver clock offset with respect to the GPS time, the receiver clock offset in each epoch is estimated by using the ground truth receiver positions provided from Ublox EVK-M8U and the precise timing information, such as the receiver clock drift and the receiver clock bias, provided from Ublox EVK-M8T. We should note that the pseudorange of the HAPS in the experiment is modeled as a function of the receiver clock offset, which is estimated with best effort. Nevertheless, additional errors should be expected in the pseudorange of the HAPS with the magnitude depending on the quality of all visible satellite signals and the ground truth receiver position. As we would expect the quality of the satellite signals in the suburban area to be better than that in the dense urban area, we should also expect the receiver clock offset to be estimated with higher accuracy in the suburban area than in the dense urban area.

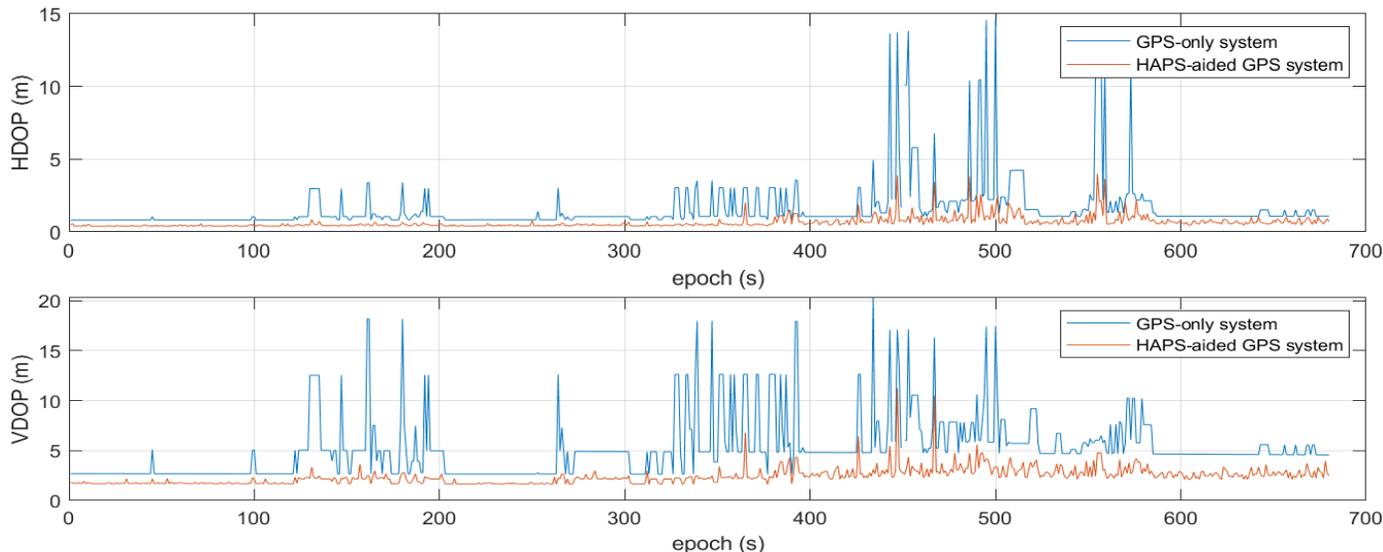

Fig. 11. HDOP (top) and VDOP (bottom).

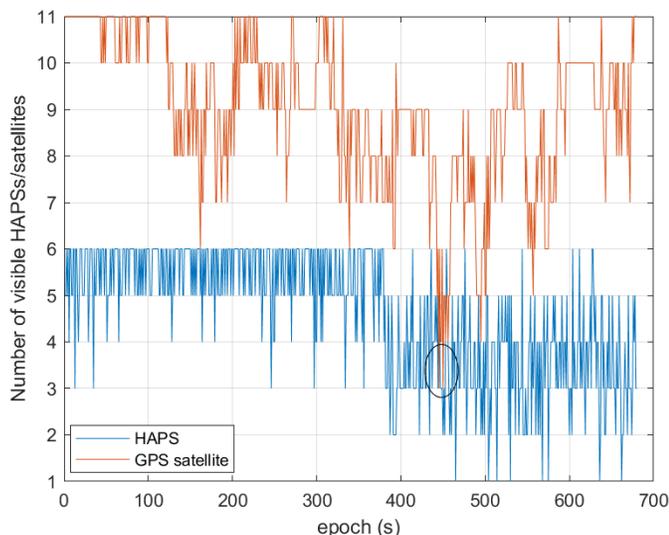

Fig. 10. HAPS availability during the entire observation period.

*B. Experiment Results*

With more ranging sources, we should expect the availability of the HAPS-aided GPS system to be higher than the GPS-only system. The availability of HAPS and GPS satellite during the entire course of observation is shown in Fig. 10. As we can see, the availability of the HAPS-aided GPS system during the entire observation period is 100 %, while the availability of the GPS-only system is 99.71 % as there are two epochs (circled in a black ellipse) where the number of GPS satellites is three. While the difference between the availability of the HAPS-aided GPS system and the GPS-only system is not significant, this is probably because Ottawa is a relatively small metro city compared to the metro cities like Chicago. In the following, we first present a comparison of the HDOP and VDOP between the HAPS-aided GPS system and the GPS-only system. Next, we analyze the 3D positioning performance for both the GPS-only system and the HAPS-aided GPS system. To show the improvements brought by the RAIM algorithm, we compare the RAIM-enabled positioning systems, where both the SPP and the RAIM algorithms are applied, with the RAIM-disabled positioning systems, where only the SPP algorithm is applied.

*1) Dilution of Precision Analysis*

Fig. 11 shows the HDOP and VDOP of the GPS-only system and the HAPS-aided GPS system. As we can see, both the HDOP and VDOP of the HAPS-aided GPS system are better than that of the GPS-only system. In particular, we notice that there are fewer spikes on the HDOP and VDOP performance of the HAPS-aided GPS system, which demonstrates that the

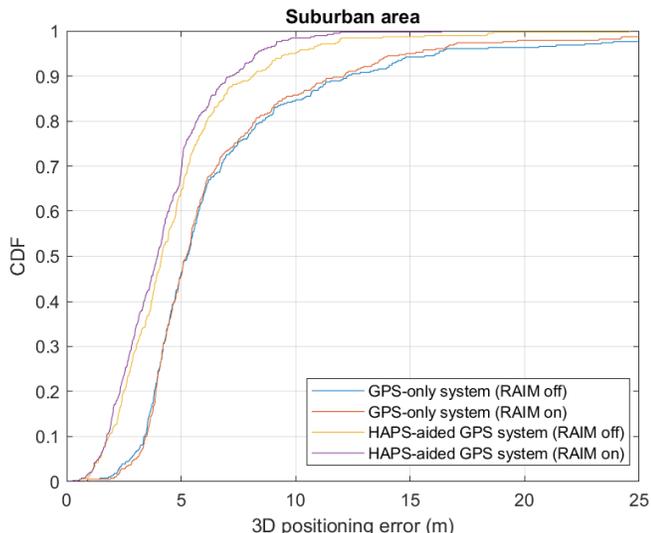

Fig. 12. CDF of the 3D positioning accuracy in the suburban area.

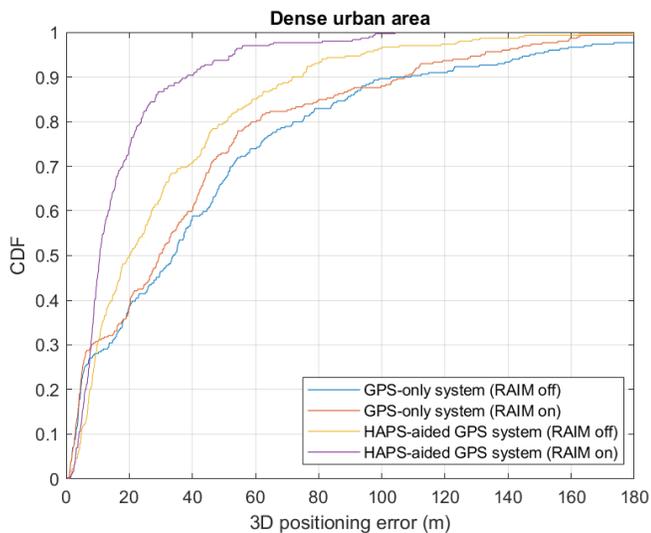

Fig. 13. CDF of the 3D positioning accuracy in the dense urban area.

HDOP and VDOP performance of the HAPS-aided GPS system is more stable than the GPS-only system.

*2) 3D Positioning Accuracy Analysis*

The CDFs of the 3D positioning accuracy for both the suburban area and the dense urban area are shown in Fig. 12 and Fig. 13, respectively. As we can see, without enabling the RAIM, the 90-percentile 3D positioning accuracy of the GPS-only system can be improved by 36 % in the suburban area, and 33.64 % in the dense urban area with the assistance from the HAPS. With the RAIM turned on, we can observe that the positioning performance of both the GPS-only system and the HAPS-aided GPS system can be further improved. Yet we notice that the improvement brought by the RAIM in the suburban area for the GPS-only system is almost negligible, while it is more tangible for the HAPS-aided GPS system. This is because the quality of signals in the suburban area tends to be great, and the HAPS-aided GPS system has a relatively higher chance to enable the RAIM as there are more ranging sources in the system. This observation is also applicable to the dense urban scenario where the 90-percentile 3D positioning accuracy of the HAPS-aided GPS system is improved by 45.2 %, which is much more significant than that for the GPS-only system. For the dense urban scenario where the multipath is severe, the RAIM algorithm plays a more significant role, especially in the HAPS-aided GPS system. The reason behind this is that the number of visible satellites in the dense urban area is low, making the RAIM algorithm for the GPS-only system less effective. Since the implemented RAIM algorithm detects and excludes an abnormal observation by multiplying its variance with an exponential term if the absolute value of its normalized pseudorange residual surpasses the critical value, we count the number of times where the absolute value of the normalized pseudorange residuals surpass the critical value for both systems considered and for both the suburban scenario and the dense urban scenario. For convenience, we rephrase the number of times where the absolute value of the normalized pseudorange residuals surpass the critical value as the number of times the RAIM is enabled. With the system model considered in this work, we find that the number of times the RAIM is enabled for the GPS-only system is roughly 23.84 % as many as the number of times the RAIM is enabled for the HAPS-aided GPS system in the suburban area; and the number of times the RAIM is enabled for the GPS-only system is about 50.72 % as many as the number of times the RAIM is enabled for the HAPS-aided GPS system in the dense urban area. This demonstrates the applicability of the RAIM algorithm on the HAPS-aided GPS system, especially in the dense urban area.

## V. Conclusion

HAPS have a number of advantages over satellites, including (but not limited to) lower latency, lower pathloss, smaller pseudorange errors, and HAPS can provide continuous coverage to reduce the number of handovers for users. This makes HAPS an excellent candidate to serve as another type of ranging source. Since urban areas are where GNSS positioning performance degrades severely, while also being where most people live, deploying several HAPS as additional ranging sources above metropolitan cities would improve GNSS positioning performance and maximize the profit of the extra payloads on HAPS. From both the simulation and physical experiment results, we observed that HAPS can indeed improve the HDOP, the VDOP, and the 3D positioning accuracy of a legacy GNSS. With the system model considered in this work, we showed that the 90-percentile 3D positioning accuracy of the GPS-only system can be improved by around 35 % in both suburban and dense urban areas. We demonstrated the applicability of the RAIM algorithm for the HAPS-aided GPS system, especially in the dense urban areas. To enhance the simulation of the HAPS-aided GPS, the receiver clock offset should be estimated with higher accuracy. We think the effectiveness of the RAIM algorithm can be improved if the standard deviation of the target observable is available. To further improve the positioning performance for urban areas, we can make use of terrestrial signals such as cellular network signals and multipath signals. This would constitute a vertical heterogeneous network (V-Het-Net) positioning system, which we believe will yield a lower VDOP based on the DOP illustration presented in this paper.

## ACKNOWLEDGMENT

The Skydel software is a formal donation from Orolia to Carleton University.

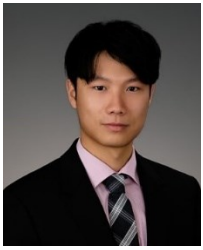

**Hongzhao Zheng** (Member, IEEE) received the B. Eng. (Hons.) degree in engineering physics from the Carleton University, Ottawa, ON, Canada, in 2019. He is currently a PhD student at Carleton University. His research interest is the urban positioning using sensor-enabled heterogeneous wireless infrastructure.

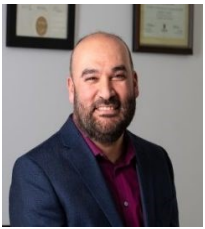

**Mohamed Atia** (Senior Member, IEEE) received the B.S. and M.Sc. degrees in computer systems from Ain Shams University, Cairo, Egypt, in 2000 and 2006, respectively, and the Ph.D. degree in electrical and computer engineering from Queen's University, Kingston, ON, Canada, in 2013. He is currently an Associate Professor with the Department of Systems and Computer Engineering, Carleton University. He is also the Founder and the Director of the Embedded and Multi-Sensory Systems Laboratory (EMSLab), Carleton University. His research interests include sensor fusion, navigation systems, artificial intelligence, and robotics.

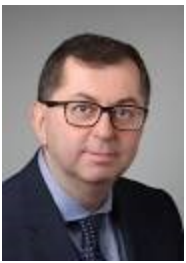

**Halim Yanikomeroglu** (Fellow, IEEE) received the BSc degree in electrical and electronics engineering from the Middle East Technical University, Ankara, Turkey, in 1990, and the MASc degree in electrical engineering (now ECE) and the PhD degree in electrical and computer engineering from the University of Toronto, Canada, in 1992 and 1998, respectively. Since 1998 he has been with the Department of Systems and Computer Engineering at Carleton University, Ottawa, Canada, where he is now a Full Professor. His research interests cover many aspects of wireless communications and networks. He has given 110+ invited seminars, keynotes, panel talks, and tutorials in the last five years. Dr. Yanikomeroglu's collaborative research with industry resulted in 39 granted patents. Dr. Yanikomeroglu is a Fellow of the IEEE, the Engineering Institute of Canada (EIC), and the Canadian Academy of Engineering (CAE). He is a Distinguished Speaker for the IEEE Communications Society and the IEEE Vehicular Technology Society, and an Expert Panelist of the Council of Canadian Academies (CCA|CAC).

Dr. Yanikomeroglu is currently serving as the Chair of the Steering Committee of IEEE's flagship wireless event, Wireless Communications and Networking Conference (WCNC). He is also a member of the IEEE ComSoc GIMS, IEEE ComSoc Conference Council, and IEEE PIMRC Steering Committee. He served as the General Chair and Technical Program Chair of several IEEE conferences. He has also served in the editorial boards of various IEEE periodicals.

Dr. Yanikomeroglu received several awards for his research, teaching, and service, including the IEEE ComSoc Fred W. Ellersick Prize (2021), IEEE VTS Stuart Meyer Memorial Award (2020), and IEEE ComSoc Wireless Communications TC Recognition Award (2018). He received best paper awards at IEEE Competition on Non-Terrestrial Networks for B5G and 6G in 2022 (grand prize), IEEE ICC 2021, IEEE WISEE 2021 and 2022.